# IMPROVED QUANTIFICATION OF NAPHTHALENE USING NON-LINEAR PARTIAL LEAST SQUARES REGRESSION

*M. Bastuck*, *M. Leidinger, T. Sauerwald, A. Schütze*

Lab for Measurement Technology, Saarland University, D-66123 Saarbrücken, Germany
m.bastuck@lmt.uni-saarland.de

## ABSTRACT

A test dataset is generated using temperature cycled operation with a $WO_3$ metal oxide semiconductor (MOS) gas sensor. Six concentrations of naphthalene from 0 to 40 ppb are measured and, subsequently, used to evaluate the performance of three variants of Partial Least Squares Regression (PLSR). Ordinary PLSR produces highly non-linear models due to the non-linear response of the sensor. Double-logarithmic data results in a model with much better linearity which has a resolution of 4 ppb in the range from 0 to 20 ppb. The more complex Locally Weighted PLSR (LW-PLSR) produces an even better model, especially for higher concentrations, without making any assumptions for relationships in the underlying data.

*Index terms*– Volatile Organic Compounds (VOC), multivariate analysis, temperature cycled operation (TCO)

## 1. INTRODUCTION

Many Volatile Organic Compounds (VOCs) like naphthalene are hazardous already at very low concentrations (ppb and sub-ppb). The short-term symptoms span a wide range and cancer can be a consequence when the exposure limit of 2 ppb is exceeded permanently [1]. Naphthalene and other VOCs are widely used, e.g. in solvents, and thus low-cost systems for identification and quantification are desirable.

Currently, no such system exists. It has been shown that temperature cycled operation (TCO) can increase sensitivity and selectivity of conventional metal oxide semiconductor (MOS) sensors drastically [2], but so far, most work has focused on discrimination rather than quantification of gases in this context.

In this work, the performance of three variants of the well-established Partial Least Squares Regression algorithm [3] on the same dataset is evaluated.

## 2. METHODS

### 2.1 Experimental setup

Tungsten trioxide ($WO_3$) was deposited by pulsed laser deposition (PLD) on a micro heater structure. TCO is used to increase the sensor's sensitivity and selectivity. The temperature cycle consists, here, of two ramps: from 200 to 400 °C in 20 s, and back again to 200 °C in 20 s. It is repeated during the whole measurement and the sensor's conductivity is recorded at 4 Hz.

The MOS sensor is exposed to ppb-level concentrations of naphthalene, provided by the gas mixing apparatus described in [4]. Concentrations were 40, 20, 10, 5 and 2.5 ppb, applied for 15 min each and followed by 15 min of pure background (humid air, 20 % r.h.), respectively. Afterwards, the same concentrations were applied again in reverse order (i.e. 40 ppb last). This enables inclusion of sensor drift in the data treatment. The test dataset contains approx. 30 cycles for each gas concentration plus 100 for background, sampled before the first gas exposure.

### 2.2 Data treatment

Each temperature cycle was divided into ten equal ranges. Mean value and slope were computed for each range resulting in 2×10 features (or, from here, simply "data"). This corresponds to a dimensionality reduction from 160 (raw data points per cycle) to 20 while keeping most of the information.

Basically, the Partial Least Squares Regression algorithm [3] is employed to build a quantification model out of the obtained features. Three different variants are compared.

#### 2.2.1 Original PLSR and validation

The algorithm published in [3] projects the features in a new, usually lower-dimensional space. This space is rotated to find the best compromise between linearity of the data and covariance to the concentration. The number of dimensions of this new space is called "components". Too few will result in loss of information and a poor model, while too many will lead to an overfitted model with poor prediction ability. In order to find the best model, the Root Mean Squared Error of Cross Validation (RMSECV) is calculated for all numbers of components using leave-one-out cross-validation (LOOCV, [5]). The model yielding the lowest RMSECV with fewest components (within some tolerance to account for fluctuations) is considered the best one.

#### 2.2.2 PLSR with data pre-treatment

The same algorithm as before is used, but with preprocessed data. The response of cyclically operated MOS sensors usually does not show a linear relationship with concentration and there is yet no universal theoretical model providing a functional relationship. For MOS sensors at static temperatures the response S scales with the concentration c usually in form of a power law $S = a \times c^b$, which has been conclusively found in numerous empirical studies and sensors models [6],[7],[8]. It seems that this approach is useful also for dynamic operation as many features appear linear in a double-logarithmic plot (not shown here). Therefore, the PLSR is computed on logarithmic features and concentrations. Adding "1 ppb" to each concentration avoids problems with zero concentration. In the final model, the data are delogarithmized and the shift is subtracted again.

#### 2.2.3 Locally Weighted PLSR

Locally Weighted PLSR (LW-PLSR) is a non-linear variant of PLSR [9]. Instead of training a model and obtaining a set of coefficients for projecting unknown data, this algorithm builds a new, local model for each new data point based on the k nearest points in feature space. The new data point is excluded from model-building and projected afterwards using the resulting coefficients. The optimal number of components is determined as described in section 2.2.1, and k is chosen so that each local model contains at least two different concentrations during LOOCV.

#### 2.2.4 Assessment of model performance

The RMSECV is a suitable measure when comparing the prediction ability of different models. A RMSECV much higher than RMSE can reveal overfitting. Instead of a correlation coefficient like Pearson's R which assumes normally distributed data, we introduce Root Mean Squared Error of Means (RMSEM) to quantify linearity. Lower values correspond to a more linear model. Uncertainty determines the resolution of a model and is here defined as $2 \times 2\sigma_{max}$, the end-to-end distance of the largest error bar in a model. It gives an idea which concentration changes can be resolved by the model.

### 3. RESULTS AND DISCUSSION

#### 3.1 Comparison of data treatment approaches

Trying to find a linear relationship of raw data and concentration using PLSR (cf. section 2.2.1) results in a distinct curvature of the model (Fig. 1). A discontinuity for 2.5 ppb, whose mean value in the model lies slightly below the one for 0 ppb, increases non-linearity even further. This is reflected in a RMSEM of over 5 ppb. RMSE and RMSECV are similar, i.e. no overfitting is present. The uncertainty, 6.3 ppb, is approx. 16 % of the maximum model concentration, but does not take deviations of the mean into account and is thus over-optimistic.

The model can be improved drastically by applying PLSR on linearized, i.e. double-logarithmic, data (cf. section 2.2.2). This could be expected since PLSR searches for linear relationships in the data. Using 15 components as before, the RMSEM decreases almost to 0.1 ppb (Fig. 1b), i.e. the model is almost perfectly linear. RMSE and RMSECV are similar and, moreover, almost 70 % lower than for the first model. This decrease must, in a large part, be accounted to increased linearity. Nevertheless, some improvement is obtained by having significantly lower uncertainties, i.e. 2.6 ppb, in the range up to 20 ppb, which is nearly 2.5 times better than first model. The increasing uncertainty for higher concentrations is due to the fact that PLSR eventually uses the least squares approach which tries to cancel out all residuals. Hence, the errors of all concentrations are roughly equal in a double-logarithmic plot (not shown), and increase exponentially with the concentration when delogarithmized. Despite the slightly higher overall uncertainty, this model is better suited for quantification than the first one due to its good linearity.

The third approach (Fig. 1c), LW-PLSR (cf. section 2.2.3), achieves equally good linearity as the second model, which can be considered validation of the linearization choice in the previous algorithm. The decrease in uncertainty to 2.5 ppb and RMSE to 0.14 ppb can be attributed to the very low uncertainty for the highest concentration in this model. For all other concentrations, the model performs only marginally better than the model using logarithmic data. The price for this improvement is a much higher computational effort which can be especially difficult to handle for low-cost or real-time systems. Hence, LW-PLSR is foremost interesting for academic purposes because it is able to extract a maximum of information, without relying on any assumptions on the functional relationship of the data.

#### 3.2 Sensor stability

MOS sensors with PLD-deposited $WO_3$ are a very recent development, thus, no extensive research regarding their long-term stability has been carried out. First results show declining sensitivity to naphthalene after some days of continuous operation. Until then, however, the PLD-MOS sensors exhibit better sensitivity to naphthalene than comparable commercial sensors, which is why they have been chosen for this work.

### 4. CONCLUSIONS AND OUTLOOK

It has been shown that the response of a $WO_3$-PLD-MOS sensor with TCO can be linearized and, subsequently, PLSR can be used for quantification of naphthalene in the ppb-range. In a limited range up to 20 ppb, an uncertainty of 2.6 ppb has been achieved. The more complex LW-PLSR algorithm performs only slightly better, but does so without any a priori assumptions about relationships in the data.

### 5. ACKNOWLEDGEMENTS

This project has received funding from the European Union's Seventh Framework Programme for research, technological development and demonstration under grant agreement No 604311.

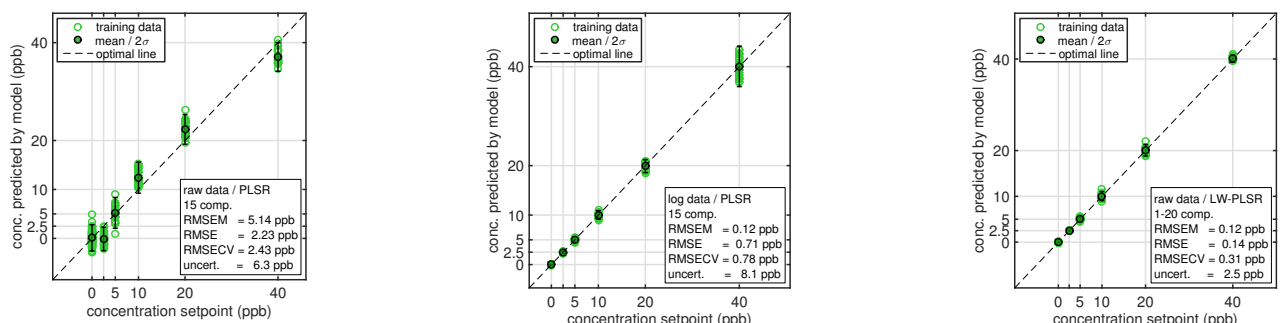

**Fig. 1. (a) PLSR with raw data, (b) PLSR with logarithmic data, and (c) LW-PLSR with raw data.**